\documentclass[12pt,a4paper]{article}
\usepackage{amsfonts}

\begin{document}

\title{Localized gravity in non-compact superstring models
\footnote{SPhT-T04/115, CPHT-PC-051.0904, hep-th/0409197}} 
\author{Emmanuel Kohlprath$\, ^a$ \textbf{and} Pierre Vanhove$\, ^b$\\
\mbox{}\vspace*{-0.3cm}\\
$^a$\textit{Centre de Physique Th{\'e}orique}\\ 
\textit{{\'E}cole Polytechnique}\\ 
\textit{91128 Palaiseau-Cedex, France}\\
{\tt {\small email: emmanuel.kohlprath@cpht.polytechnique.fr}}\\
$^b$\textit{Service de Physique Th{\'e}orique, CEA/DSM/PhT}\\
 \textit{Unit{\'e} de recherche associ{\'e}e au CNRS}\\
\textit{CEA/Saclay, 91191 Gif-sur-Yvette, France}\\
{\tt {\small email: pierre.vanhove@cea.fr}}
\mbox{}\vspace*{-0.3cm}}

\date{}

\maketitle

{\bf Abstract}
{
We discuss a string-theory-derived mechanism for localized gravity,
which produces a deviation from Newton's law of gravitation at
cosmological distances. This mechanism can be realized for general
non-compact Calabi-Yau manifolds, orbifolds and orientifolds. After
discussing the cross-over scale and the thickness in these models we
show that the localized higher derivative terms can be safely
neglected at observable distances. We conclude by some observations on
the massless open string spectrum for the orientifold models.
}

\section{Introduction}

Extra dimensions are a natural concept in string theory which brings
many new options on how to think about gravity couplings in our
world. Interestingly, when the extra dimensions are non-compact the
dilution of the 
gravitational interactions  
into the bulk affects the effective four-dimensional potential over
cosmological scale.  
This was first pointed out in the so-called 
{\sc dgp} model~\cite{DGP} where  gravity is quasi-localized in four
dimensions. Remarkably, this model has late time self-expanding 
cosmological solutions, which do not need the introduction of a
cosmological
constant~\cite{Deffayet:2000uy,Deffayet:2001pu,Dvali:2002fz}.  
The main properties of the {\sc dgp} model with one or more
non-compact extra-dimensions have been reviewed in 
\cite{GigaReview}.
In the following we discuss a string theoretical setup of induced
gravity.  
We first consider the issue of localizing  gravitational interactions
following the references~\cite{AMV, Kohlprath:2003pu} (see \cite{KTT}
for a different realization and \cite{Kiritsisreview1} for a review)
and then address the question about  constructing a consistent 
gauge theory sector within the model of localized  gravity.

\section{The induced gravity  model}

The {\sc dgp}  model and its generalization are specified by a bulk
Einstein-Hilbert ({\sc eh}) term and a four-dimensional term
\begin{equation}
M^{2+n} \int_{M_{4+n}} d^4x d^ny\, \sqrt{|G|}\, \mathcal{R}_{(4+n)} + M_{pl}^2
\int_{M_4} d^4x\sqrt{|g|}\, \mathcal{R}_{(4)}\, ,
\label{e:dgp}
\end{equation}
with $M$ and $M_{pl} (=:\sqrt{r_c^n M^{2+n}})$ the  (possibly
independent) respective Planck 
scales. The scale $M \geq 1$~TeV would be related to the short-distance scale
below which {\sc uv} quantum gravity or stringy effects are seen. $M_{pl}\sim
{10}^{19}$~GeV is our four-dimensional Planck mass. The four-dimensional
metric is the restriction of the bulk metric
$g_{\mu \nu}=\left.G_{\mu \nu}\right|$ and we assume the {\sc
world}\footnote{We avoid calling $M_4$ a brane, since gravity
localizes on singularities of  orbifold fixed points \cite{AMV},
orientifold planes \cite{Kohlprath:2003pu}, and 
intersection of branes \cite{Epple:2004ra}. In all these mechanisms,
four-dimensional gravity is induced by  loops of 
localized twisted fields coupled to the background metric. In a
sense these mechanisms are string theory realizations of  
the field theory scenario reviewed in \cite{Adler:1982ri}.}
rigid, allowing the gauge
$\left. G_{i\mu }\right|=0$ with $i\geq 5$. Finally no extrinsic
curvature terms (as the Gibbons--Hawking term) are needed.

The effective potential between two test masses in four dimensions \cite{GS}
\begin{eqnarray}
\label{e:Force}
\int d^3x \, e^{-i p\cdot x} \, V(x)&=&\frac{D(p)}{1+ r_c^n\, p^2 \,
D(p)}  \, \left[\tilde T_{\mu \nu} 
T^{\mu \nu}- g(p) \, \tilde T^\mu _\mu \, T^\nu_\nu\right]\\
D(p) &=& \int d^nq \, \frac{f_w(q)}{p^2+q^2}\label{e:Dp}\\
g(p) &=& \frac{1}{2}\Bigr[ \frac{(2-2n)p^2D(p)-2/r_c^n}
{(2-2n)p^2D(p)-(2+n)/r_c^n}\Bigl]
\end{eqnarray}
is a function of the bulk graviton retarded Green's function
$G(x,0;0,0)=\int d^4p \, e^{i p\cdot x}\, D(p)$ evaluated for two
points localized on the {\sc world} ($y=y'=0$).  The
integral~(\ref{e:Dp}) is {\sc uv}-divergent for $n>1$ unless there is
a non-trivial brane thickness profile $f_w(q)$ of width $w$. If the
four-dimensional {\sc world} has zero thickness, $f_w(q)\sim 1$, the bulk
graviton does not have a normalizable wave function. It therefore
cannot contribute 
to the induced potential, which always takes the form  $V(p) \sim 1/p^2$ and
Newton's law remains four-dimensional at all distances.  For a non-zero
thickness $w$, there is only one crossover length scale, $R_c$: 
\begin{equation}
\label{e:Rc}
R_c=w\left(\frac{r_c}{w}\right)^{\frac{n}{2}}\, , 
\end{equation}
above which one obtains a higher-dimensional
behaviour~\cite{KTT}. Therefore the effective
potential presents two regimes: (i) at short distances ($w\ll r\ll R_c$) the
gravitational interactions are mediated by the localized four-dimensional
graviton and Newton's potential on the {\sc world} is given by $V(r)\sim
1/r$ and, (ii) at large distances ($r\gg R_c$) the modes of the bulk graviton
dominate, changing the potential. For $n=1$ the expressions~(\ref{e:Force})
and~(\ref{e:Dp}) are finite and unambiguously give $V(r)\sim 1/r$ for
$r\gg r_c$.\footnote{ For $n=1$ the
propagator~(\ref{e:Dp}) is not {\sc uv}-divergent, but~(\ref{e:Rc})
predicts a critical radius $R_c=\sqrt{w r_c}\ll r_c$ below
which graviton's Kaluza--Klein excitations (induced by the cutoff) become
massless, and the theory is five-dimensional. See \cite{GigaReview}
for a lucid discussion of the perturbation theory for the 5d model.}
For a co-dimension bigger than 1, the precise behaviour for 
large-distance interactions depends \emph{crucially} on the {\sc uv}
completion of the theory. Embedding this scenario in string theory
will allow us to derive unambiguously 
all the physical parameters of the model.

At this point we stress a fundamental difference with the \emph{finite extra
dimensions} scenarios. In these cases Newton's law gets higher-dimensional
at distances smaller than the characteristic size of the extra dimensions.

\section{String Theory realization}
We explain following~\cite{AMV} how to realize~(\ref{e:dgp}) with $n=6$
as the low-energy effective action of string theory on a non-compact
six-dimensional manifold $\mathcal{M}_6$. We work in the context of
$\mathcal{N}=1$ and
$\mathcal{N}=2$ supergravities in four dimensions but the mechanism for
localizing gravity is independent of the number of supersymmetries. Of course
for  $\mathcal{N}\geq 3$ supersymmetries, there is no localization. 

In string perturbation, corrections to the four-dimensional Planck mass are
in general very restrictive. In the heterotic string, they vanish to
all orders in perturbation theory~\cite{AGN}; in type {\sc i} theory,
there are moduli-dependent corrections generated by open
strings~\cite{AB}, but they vanish when the
manifold $\mathcal{M}_6$ is decompactified; in type {\sc ii}
theories, they are constant, 
independent of the moduli of the manifold $\mathcal{M}_6$, and receive
contributions only from tree and one-loop levels (at least for
supersymmetric backgrounds)~\cite{AMV,KiritsisKounnas95,AFMN}. 

The origin of the two {\sc eh} terms in~(\ref{e:dgp}) can be traced
back to the perturbative corrections to the eight-derivative effective 
action of type~{\sc II} strings in ten dimensions.  These corrections
include the tree-level and one-loop terms given 
by:\footnote{The rank-eight tensor $t_8$ is defined as
$t_8M^4\equiv -6({\rm tr}M^2)^2 +24{\rm tr}M^4$, and the $\pm $ sign
depends on the chirality (type~{\sc iia/b}) of the
theory. See~\cite{Peeters:2000qj} for more details.}
\begin{eqnarray}
\label{e:Action}
&&\frac{M_s^8}{(2\pi)^7}\int\limits_{M_{10}}\!\!d^{10}x\sqrt{|G|}\,
\frac{1}{g_s^2} \mathcal{R}_{(10)} + 
\frac{M_s^2}{3(4\pi)^7}\int\limits_{M_{10}} \!\!d^{10}x\sqrt{|G|}\,  \left(
\frac{2\zeta(3)}{g_s^2} +  4\zeta(2)\right)t_8t_8 R^4 \nonumber \\
&-&\frac{M_s^2}{3(4\pi)^7} \int\limits_{M_{10}} \left(
\frac{2\zeta(3)}{g_s^2} 
\mp 4\zeta(2)\right) R\wedge R \wedge R\wedge R \wedge e \wedge e
+\cdots
\end{eqnarray}
where $M_s$ is the string scale and  $\phi$ is the dilaton
field determining the string coupling $g_s=e^{\langle\phi\rangle}$.

On a direct product space-time $\mathcal{M}_6 \times \mathbb{R}^4$ the
term $t_8t_8 R^4$ 
contributes in four dimensions to $R^2$ and $R^4$ terms~\cite{AFMN} (and
to a cosmological constant which is zero due to $\mathcal{N}=2$
supersymmetry~\cite{AMV}).  At the level of 
zero modes the second $R^4$ term splits as $\frac{1}{3!(4\pi)^3}
\int_{M_6} \, R\wedge R \wedge R \times 
\int_{M_4} \, \mathcal{R}_{(4)}=\chi\, \int_{M_4}\,
\mathcal{R}_{(4)}$, and we have

\begin{equation}
\label{e:Local}
\frac{M_s^8}{(2\pi)^7}\int\limits_{M_4 \times M_6}\!\!d^{10}x
\sqrt{|G|} \, \frac{1}{g_s^2}\, \mathcal{R}_{(10)} +  
\frac{\chi M_s^2}{(2\pi)^4} \int\limits_{M_4} \!\!d^{4}x
\sqrt{|g|}\, \left(-\frac{2\zeta(3)}{g_s^2} 
\pm   4\zeta(2)\right) \mathcal{R}_{(4)}\, ,
\end{equation}
which gives the expressions for the Planck masses $M$ and $M_{pl}$ for
type {\sc ii}.  A number of 
conclusions (confirmed by string calculations
in~\cite{AMV,Kohlprath:2003pu,Epple:2004ra}) can be 
reached by looking closely at~(\ref{e:Local}):

$\triangleright$  $M_{pl}\gg M$ requires a large non-zero Euler
characteristic for $M_6$ and/or a weak string coupling constant
$g_s\to0$ ($M_s g_s^{-1/4}$ gives the scale of the $\mathcal{R}_{(10)}$ 
term and $M_s g_s^{-1}$ the scale of the tree level $\mathcal{R}_{(4)}$ term).

$\triangleright$ Since $\chi$ is a topological invariant the localized
$\mathcal{R}_{(4)}$ term coming from the closed string sector is universal,
independent of the background geometry and dependent only on the
internal topology.\footnote{In type 
{\sc iia/b}, $\chi$ counts the difference between the numbers of
$\mathcal{N}=2$ vector multiplets and hypermultiplets: $\chi=\mp 2(n_V-n_H)$
(where the graviton multiplet counts as one vector). Field theory
computations of~\cite{Adler:1982ri} show that the Planck mass
renormalization depends on the {\sc uv} behaviour of the matter fields
coupling to the   external metric. But, even in the
supersymmetric case, the corrections are not obviously given by an
index.}  It is a matter of 
simple inspection to see that if one wants to have a localized {\sc eh} term
in less than ten dimensions, namely something linear in curvature, with
non-compact internal space in all directions, \emph{the only possible
dimension is four} (or five in the strong coupling limit but this
setup does not have an interesting phenomenology). 

$\triangleright$ The width is given by the four-dimensional induced
Planck mass \cite{AMV}
\begin{equation}
\omega \simeq M_{pl}^{-1}.
\end{equation}

\section{Orbifold and Orientifold models}

The supersymmetric $\mathbb{Z}_N$ orbifolds or orientifolds do not
have a tree level  induced {\sc eh} term, but one-loop
contributions to the induced Einstein term from the torus
$\mathcal{T}$, the annulus $\mathcal{A}$, the Moebius strip
$\mathcal{M}$ and the Klein bottle $\mathcal{K}$. We present the
arguments of~\cite{AMV,Kohlprath:2003pu} 
that show that the (quasi-)localization of gravity is purely a closed
string sector phenomenon and that the open string sector
contributions from $\mathcal{A}$, $\mathcal{M}$ and $\mathcal{K}$ are
always subleading or vanishing. Therefore orbifold and orientifold
models give the same estimates for the width and for the crossover
scale.  

\medskip
$\triangleright$ The closed string one-loop graviton amplitudes (from
the torus) take the form of sums of quasi-localized 
contributions at the positions of the fixed points $x_f$~\cite{AMV}.
Focusing on one particular fixed point $x_f=0$ and sending the radii to
infinity, we obtain the effective action for the quasi-localized {\sc
eh} term 
\begin{equation}
\frac{\chi M_s^2}{24\pi^2} \, \int d^4x d^6y \sqrt{g} f_w(y) \,
\mathcal{R}_{(4)}\, 
\label{orbifoldresult}
\end{equation}
where $\delta M_{pl}^2=M^2_{s}\times \mathcal{O}(N)$ as
$N\rightarrow\infty$. For odd $N$ orientifold models 
the torus contribution is given by one half  
of the orbifold result (\ref{orbifoldresult}) and is $\mathcal{O}(N)$. 
For a more general non-compact background, the Euler number can be
distributed over 
the various fixed points of the  internal space, giving rise to
different localized terms, with a different value for the induced
Planck mass.\footnote{For instance, keeping two fixed points we
obtained the bi-gravity scenario discussed in \cite{Padilla:2004uj},  
with (possibly) different Planck mass at each fixed point depending
on the distribution of the twisted fields in the model.} 

$\triangleright$  The open string sector given by the sum of the
contributions $\mathcal{A}+\mathcal{M}+\mathcal{K}$ 
 is always subleading as compared with the torus contribution since
$\delta M_{pl}^2\sim M_{s}^2\times\mathcal{O}(1)$ for large-$N$  
 (actually it even vanishes for orientifold models with odd $N$ that
have no $\mathcal{N}=2$ sectors). 
 Importantly the twisted tadpole cancellation conditions imply that
the open string sector contribution to the induced Planck mass is  
  ultraviolet finite, and no new scale can arise in these models.

$\triangleright$ As pointed out in \cite{KTT} higher derivative $R^2$
terms both on the {\sc world} and in the bulk can drastically change
the picture as they introduce new scales. The induced $R^2$ terms can
be determined similarly to the induced Einstein term by considering
the piece in forth order in momentum of graviton amplitudes. The
crucial observation is that the torus contribution is again of
$\mathcal{O}(N)$ 
and the sum of the contributions of annulus, Moebius strip and Klein
bottle is $\mathcal{O}(1)$ and therefore subleading as compared to the torus
contribution. The leading contribution to the terms in the effective
action is then 
\begin{equation}
\Delta\mathcal{L}_{\textrm{\small{eff}}}^{(4)}=N M_s^2 b
\sqrt{|g|}\mathcal{R}_{(4)}+ N c_1 
\sqrt{|g|}R^2 + N c_2 \sqrt{|g|}R_{\mu\nu}R^{\mu\nu}+ N c_3
\sqrt{|g|}R_{\mu\nu\rho\sigma}R^{\mu\nu\rho\sigma}
\end{equation}
with numbers $b,c_i$, $i=1,2,3$. The induced $R^2$ terms are
negligible for $p^2\, M_{s}^{-2}\ll b/c_i$. Gravity is only measured
above 1~mm so even if $M_s$ is as low as 1~TeV the induced $R^2$ terms
can be neglegted if 
\begin{equation}
\frac{b}{c_i}\gg10^{-32}. \label{bvsc}
\end{equation}
E.g. if one computes the forth order piece of the off-shell two
gravitons amplitude similarly to \cite{Kohlprath:2003pu} one can
determine the non-ambiguous coefficient $c_3$ (the value  
of $c_1$ and $c_2$ are affected by field redefinitions). Though the
authors did not evaluate the final 
world sheet integral explicitly the expression is similar to the one
for the 
coefficient $b$ and (\ref{bvsc}) is satisfied. It is generic that
$\mathcal{O}(c_i)=\mathcal{O}(b)$ as long as the corresponding term is
not protected by some symmetry, and the hierarchy is controlled by the
value of $M_s$.  
The discussion generalizes for all induced higher
derivative terms. The conclusion is that the induced higher derivative
terms in the orbifold \cite{AMV} and orientifold models
\cite{Kohlprath:2003pu} of induced gravity can be neglegted at
observable distances. We expect this to be valid for general
non-compact Calabi-Yau, too. In contrast to them higher derivative
terms in the bulk need further study (see \cite{KTT} for instance).
\section{Phenomelogical implications}

The crossover radius of eq.~(\ref{e:Rc}) is given by the string
parameters (for $n=6$) 

\begin{equation}
\label{e:RcS}
R_c = \frac{r_c^3}{w^2}\simeq (2\pi)^{7/2} g_s\frac{M_{pl}^3}{M_s^4}
\simeq g_s\times 10^{34}\ {\rm cm}\, , 
\end{equation}
for $M^8=M_s^8/((2\pi)^7 g_s^2)$ and $M_s\simeq 1$ TeV.  Because $R_c$
has to be of cosmological scale, the 
string coupling can be relatively small, and $|\chi|\simeq 10^3 g_s^2\,
M_{pl}^2/M_s^2 \sim g_s^2\times 10^{35}$ must be very large. The
hierarchy is obtained mainly thanks to 
the large value of $\chi$, so that lowering the bound on $R_c$ lowers
the value of $\chi$.  Our actual knowledge
 of gravity at very large distances
indicates~\cite{Lue} that $R_c$ should be of the order of the Hubble
radius $R_c \simeq 10^{28}$ cm, which implies $g_s \geq 10^{-6}$ and
$|\chi|\geq 10^{23}$.  A large Euler number implies only a large
number of closed string 
massless particles. All these particles are localized
at the fixed points and should have sufficiently suppressed
gravitational-type couplings, so that their presence with such a huge
multiplicity does not contradict observations. In orbifold models we
can for instance introduce the observable gauge and matter sectors on
D3-branes placed at the position where gravity localization occurs and 
they are otherwise unconstrained. In orientifold models we already
have an open sector and we will determine the massless open string
spectrum for some examples in the next section.
Note that these results depend
crucially on the scaling of the width $w$ in terms of the Planck length:
$w\sim M_{pl}^{-\nu}$, implies $R_c\sim M_{pl}^{2\nu +1}$ in string
units. If there 
are models with $\nu>1$, the required value of $\chi$ will be much lower,
becoming $\mathcal{O}(1)$ for $\nu\ge 3/2$. In this case, the hierarchy will
be determined by tuning the string coupling to infinitesimal values,
$g_s\sim 10^{-16}$.

\section{Localization of gauge interactions}

After having discussed how gravity is localized in non-compact
orbifold and orientifold  
models, we now  discuss the localization of
gauge interactions.  We saw that the hierarchy between the bulk and
induced Planck mass required a huge number of twisted fields
(i.e. $N$ has to be large). We analyze the open spectrum of 
the non-compact $\mathbb{Z}_{N}$ orientifolds constructed in
\cite{Kohlprath:2003pu} in order to determine if a consistent 
gauge theory sector can be induced together with gravity.

\medskip

Let us consider the supersymmetric non-compact type {\sc iib}
$\mathbb{Z}_{N}$ orientifold models constructed in
\cite{Kohlprath:2003pu} and 
defined by the combined action $\Omega J$ of the worldsheet
parity transformation $\Omega$ and  
\begin{equation}
Z^{i}\to e^{2i\pi v_i}\, Z_{i}; \quad JZ^{i}=-Z^{i};\quad 
\sum_{i=1}^{3} \, v_{i}=0 \ ,
\end{equation}
where $Z^{i}:= X^{2i+2}+i \, X^{2i+3}$ for $i=1,2,3$. The last condition
ensure that the model is supersymmetric. 
We assume $N$ odd, therefore only D3-branes are needed. We assume
that they are sitting on 
top of the orientifold $O3_{+}$-planes.  Since we consider non-compact
orientifolds the $\mathbb{Z}_{N}$ orbifold action need not act
cristallographically and we do not need to impose the untwisted
tadpole cancellation condition. The Chan-Paton implementation matrices 
for the $n_3$ D3-branes will be denoted by the $n_3\times n_3$ matrices
$\gamma_{k,3}=\gamma_{1,3}^k$, $k=1,\dots,N-1$. 

\medskip

The massless open string spectrum in the non-compact
$\mathbb{Z}_N$ orientifold models of \cite{Kohlprath:2003pu} can be
determined using the method of \cite{Ibanez98,IbanezRabadanUranga98}.
The twisted tadpole cancellation conditions on the Chan-Paton
implementation matrices are
\begin{eqnarray}
\textrm{Tr }\gamma_{2k,3}&=&\pm 4\prod_{i=1}^3 \frac{1}{2\cos(\pi
kv_i)}\\
&=&\pm 4\prod_{i=1}^3 \frac{1}{1+e^{2\pi kv_i}},\quad
k=1,\dots,N-1  
\end{eqnarray}
with the positive sign for the $SO$ projection and the negative sign
for the $Sp$ projection. Let
us assume that $N$ is a prime number and that $v=(1/N,1/N,-2/N)$. We
define $\alpha:=e^{2\pi i/N}$ and use
$-1=\sum\limits_{k=1}^{N-1}\alpha^{k\lambda},\ \lambda=1,\dots,N-1$ to 
find
\begin{equation}
\textrm{Tr }\gamma_{2k,3}=\pm 4\left(\frac{1}{1+\alpha^k}\right)^2
\frac{1}{1+\alpha^{(N-2)k}}=\mp 4\left(\sum_{j=1}^{\frac{N-1}{2}}
\alpha^{(2j-1)k}\right)^2 \sum_{j=1}^{\frac{N-1}{2}}
\alpha^{(2j-1)(N-2)k}.
\end{equation}
As $\gamma_{1,3}=\gamma_{1,3}^{N+1}=\gamma_{2\frac{N+1}{2},3}$ we
arrive at
\begin{equation}
\textrm{Tr }\gamma_{1,3}=\mp \sum_{j_1,j_2,j_3=1}^{\frac{N-1}{2}}
\alpha^{j_1+j_2+j_3(N-2)}.
\end{equation}
There are two cases to distinguish: A) $\frac{N-1}{4}\in\mathbb{Z}$
and B) $\frac{N+1}{4}\in\mathbb{Z}$. We get
\begin{eqnarray}
&&A)\ \textrm{Tr }\gamma_{1,3}=\mp 4
\Bigl(\frac{N-1}{4}+\frac{N-1}{4}(\alpha^{-2}+\alpha^2)\nonumber\\
&&\hspace*{2.5cm}+\sum_{j=1}^{\frac{N-1}{4}-1}\Bigl(\frac{N-1}{4}-j\Bigr)
(\alpha^{-4j}+\alpha^{4j}+\alpha^{-(4j+2)}+\alpha^{4j+2})\Bigr)\nonumber\\
&&B)\ \textrm{Tr }\gamma_{1,3}=\mp 4
\left(\frac{N+1}{4}+\sum_{j=1}^{\frac{N+1}{4}-1}\Bigl(\frac{N+1}{4}-j\Bigr)
(\alpha^{-(4j-2)}+\alpha^{4j-2}+\alpha^{-4j}+\alpha^{4j})\right)\nonumber\\
\end{eqnarray}
so that $\gamma_{1,3}$ is a $n_{3}\times n_{3}$-block diagonal matrix
reading 
\begin{eqnarray}
&&A)\ \gamma_{1,3}=\textrm{diag}(I_{M\mp4\frac{N-1}{4}}, \alpha^{-2}
I_{M\mp4\frac{N-1}{4}}, \alpha^{2} I_{M\mp4\frac{N-1}{4}},\dots, \nonumber\\
&&\hspace*{1cm}
\alpha^{-4j} I_{M\mp4\left(\frac{N-1}{4}-j\right)}, \alpha^{4j}
I_{M\mp4\left(\frac{N-1}{4}-j\right)}, \alpha^{-(4j+2)} 
I_{M\mp4\left(\frac{N-1}{4}-j\right)}, \alpha^{4j+2}
I_{M\mp4\left(\frac{N-1}{4}-j\right)},\nonumber\\&&\hspace*{1cm}
\dots,\alpha^{-(N-1)}I_M, \alpha^{N-1}I_M) \nonumber\\
&&B)\ \gamma_{1,3}=\textrm{diag}(I_{M\mp4\frac{N+1}{4}}, \dots,
\alpha^{-(4j-2)} I_{M\mp4\left(\frac{N+1}{4}-j\right)}, \alpha^{4j-2}
I_{M\mp4\left(\frac{N+1}{4}-j\right)}, \nonumber\\
&&\hspace*{1cm} \alpha^{-4j} 
I_{M\mp4\left(\frac{N+1}{4}-j\right)}, \alpha^{4j}
I_{M\mp4\left(\frac{N+1}{4}-j\right)},\dots, \alpha^{-(N-1)}I_M,
\alpha^{N-1}I_M),
\end{eqnarray}
where $M\in\mathbb{N}$ with $M\geq \pm(N-1)$ for A) and $M\geq
\pm(N+1)$ for B). For both A) and B) this gives
\begin{equation}
n_3=NM\mp\frac{1}{2}(N^2-1)\ .
\end{equation} 
The freedom of choosing $M$ comes from
the fact that we have a non-compact model and do not need to impose
the untwisted tadpole cancellation condition.

\medskip
Let us consider the SO
projection (the Sp projection works the same). Similarly to
\cite{IbanezRabadanUranga98} we find the gauge groups 
\begin{eqnarray}
&A)& SO\Bigl(M-4\frac{N-1}{4}\Bigr) \times
U\Bigl(M-4\frac{N-1}{4}\Bigr) \times\nonumber\\
&&\ \times
\prod_{j=1}^{\frac{N-5}{4}} 
\Bigl(U\Bigl(M-4\Bigl(\frac{N-1}{4}-j\Bigr)\Bigr)\Bigr)^2 \times
U(M)\label{gaugegroup} \\
&B)& SO\Bigl(M-4\frac{N+1}{4}\Bigr) \times
\prod_{j=1}^{\frac{N-3}{4}} 
\Bigl(U\Bigl(M-4\Bigl(\frac{N+1}{4}-j\Bigr)\Bigr)\Bigr)^2 \times
U(M)\nonumber
\end{eqnarray}
which  we can write as $SO(n_0)\times
\prod\limits_{j=1}^{\frac{N-1}{2}}U(n_j)$ and the chiral spectrum
\newlength{\bax}
\newlength{\bix}
\settowidth{\bax}{$\Box$}
\settoheight{\bix}{$\Box$}
\begin{equation}
2\left(\sum_{j=1}^{\frac{N-1}{2}}(\Box_{j-1},\bar\Box_j)+
\Box\textrm{\hspace*{-\bax}\raisebox{\bix}{$\Box$}}_{\frac{N-1}{2}}
\right) +\left( (\bar\Box_{\frac{N-1}{2}-1},\bar\Box_{\frac{N-1}{2}})
+ \Box\textrm{\hspace*{-\bax}\raisebox{\bix}{$\Box$}}_1 +
(\Box_{2},\Box_{0}) + \sum_{j=1}^{\frac{N-1}{2}-3}
(\Box_{j+2},\bar\Box_j) \right).
\end{equation}
One can check that the model is  free of non-Abelian and $U(1)^3$
anomalies. Calling $n_v$ the number of $\mathcal{N}=1$ vector
multiplets and $n_{ch}$ the number of $\mathcal{N}=1$ chiral
multiplets we get
\begin{eqnarray}
&&n_v= c_{v} +\frac{4N}{3}+\frac{N^3}{6}+\frac{M^2N}{2}-
\frac{MN^2}{2}\\ 
&&n_{ch}= c_{ch} -5N +\frac{N^3}{2}+\frac{3M^2N}{2}-
\frac{3MN^2}{2},
\end{eqnarray}
where $c_v=-3/2,\ c_{ch}=9/2$ for the case A) and $c_v=3/2,\
c_{ch}=3/2$ for the case B). 
Notice that for any choice $M$ in (\ref{gaugegroup}) the number 
of $U(n_j)$ subgroups is $\mathcal{O}(N)$ and some of the $n_j$ themself are
$\mathcal{O}(N)$ as $N\rightarrow\infty$. It does therefore not seem
natural to get the standard model gauge group in these models. Notice
that this was derived for the choice that $N$ is prime and
$v=(1/N,1/N,-2/N)$. For $N$ prime and $N-1\in 4\mathbb{Z}$
with $v_1=v_2=-v_3/2=\frac{N-1}{4N}$ or $N+1\in 4\mathbb{Z}$
with $v_1=v_2=-v_3/2=\frac{N+1}{4N}$ one derives similar results and
the conclusions are the same.

\section{Conclusions}
We have discussed a mechanism in string theory to realize induced
gravity. This can be applied to non-compact Calabi-Yau manifolds,
orbifolds and orientifolds. The hierarchy between the ten-dimensional
and the four-dimensional Einstein term is due to a large Euler number
and/or weak string coupling. The thickness is given by the induced
Planck mass. The induced higher derivative terms are negligible at
observable distances and we get four-dimensional Einstein gravity
between 1~mm and the Hubble scale. The standard model can be realized
on the {\sc world} where gravity is localized in a multitude of
ways. However, the easiest orientifold realizations (though one can
easily include the standard model) seem to give way to large gauge
groups and too many chiral fields as to find these models natural and
therefore of immediate phenomenological interest. One will have to
consider more general models that may or may not be orientifold
models. We leave this for future work.

\section*{Acknowledgements}
We thank  I. Antoniadis, C. Deffayet, G. Gabadadze and E. Kiritsis for 
useful discussions. The work reported here has
been supported in part by the European Commission under the RTN contract
HPRN-CT-2000-00148 and EC Excellence Grant
MEXT-CT-2003-509661. E.K. has been supported by the
Austrian FWF project No. J2259-N02. P.V. thanks the organizers of
Quarks 2004 for the invitation to present this work. 


\begin{thebibliography}{99}

\bibitem{DGP} G.R.~Dvali, G.~Gabadadze and M.~Porrati, 
Phys.\ Lett.\ B {\bf 485} (2000) 208 [arXiv:hep-th/0005016].


\bibitem{Deffayet:2000uy}C.~Deffayet,
Phys.\ Lett.\ B {\bf 502} (2001) 199
[arXiv:hep-th/0010186].

\bibitem{Deffayet:2001pu}
C.~Deffayet, G.~R.~Dvali and G.~Gabadadze,
Phys.\ Rev.\ D {\bf 65} (2002) 044023
[arXiv:astro-ph/0105068].

\bibitem{Dvali:2002fz}
G.~R.~Dvali and G.~Gabadadze,
Phys.\ Rev.\ D {\bf 63} (2001) 065007
[arXiv:hep-th/0008054];
G.~R.~Dvali, G.~Gabadadze, M.~Kolanovic and F.~Nitti,
Phys.\ Rev.\ D {\bf 64} (2001) 084004
[arXiv:hep-ph/0102216];
G.~Dvali, G.~Gabadadze and M.~Shifman,
arXiv:hep-th/0208096;
G.~Dvali, G.~Gabadadze and M.~Shifman,
Phys.\ Rev.\ D {\bf 67} (2003) 044020
[arXiv:hep-th/0202174].

\bibitem{GigaReview}
G.~Gabadadze,
arXiv:hep-th/0408118.


\bibitem{AMV}
I.~Antoniadis, R.~Minasian and P.~Vanhove,
Nucl.\ Phys.\ B {\bf 648} (2003) 69
[arXiv:hep-th/0209030].

\bibitem{Kohlprath:2003pu}
E.~Kohlprath,
Nucl.\ Phys.\ B {\bf 697} (2004) 243
[arXiv:hep-th/0311251].

\bibitem{KTT} E.~Kiritsis, N.~Tetradis and T.~N.~Tomaras,
JHEP {\bf 0108} (2001) 012
[arXiv:hep-th/0106050].

\bibitem{Kiritsisreview1} E. Kiritsis, Fortsch. Phys. \textbf{52} (2004)
568, [arXiv:hep-th/0310001]. 

\bibitem{Epple:2004ra}
F.~Epple,
arXiv:hep-th/0408105.

\bibitem{Adler:1982ri}
S.~L.~Adler,
Rev.\ Mod.\ Phys.\  {\bf 54} (1982) 729
[Erratum-ibid.\  {\bf 55} (1983) 837].

\bibitem{GS}
G. Gabadadze and M. Shifman,
Phys. Rev. D {\bf 69} (2004) 124032 [hep-th/0312289]


\bibitem{AGN} I.~Antoniadis, E.~Gava and K.~S.~Narain, 
Phys.\ Lett.\ B {\bf 283} (1992) 209
[arXiv:hep-th/9203071].

\bibitem{AB} I.~Antoniadis, C.~Bachas, C.~Fabre, H.~Partouche and
\mbox{T.~R.~Taylor},  
Nucl.\ Phys.\ B {\bf 489} (1997) 160 [arXiv:hep-th/9608012].\break
I.~Antoniadis, H.~Partouche and T.~R.~Taylor, 
Nucl.\ Phys.\ B {\bf 499} (1997) 29 [arXiv:hep-th/9703076]. 

\bibitem{KiritsisKounnas95} E. Kiritsis and C. Kounnas,
Nucl. Phys. B \textbf{442} (1995) 472 [arXiv:hep-th/9501020].

\bibitem{AFMN} I.~Antoniadis, S.~Ferrara, R.~Minasian
and K.~S.~Narain,  
 Nucl.\ Phys.\ B {\bf 507} (1997) 571
[arXiv:hep-th/9707013]. 

\bibitem{Peeters:2000qj} K.~Peeters, P.~Vanhove and A.~Westerberg,
Class.\ Quant.\ Grav.\  {\bf 18} (2001) 843
[arXiv:hep-th/0010167].

\bibitem{Padilla:2004uj}
A.~Padilla,
Class.\ Quant.\ Grav.\  {\bf 21} (2004) 2899
[arXiv:hep-th/0402079].

\bibitem{Lue} A.~Lue and G.~Starkman,
Phys.\ Rev.\ D {\bf 67} (2003) 064002
[arXiv:astro-ph/0212083].


\bibitem{Ibanez98} G. Aldazabal, A. Font, L.E. Ibanez und G. Violero,
Nucl. Phys. B \textbf{536} (1998) 29 [arXiv:hep-th/9804026].

\bibitem{IbanezRabadanUranga98} L.E. Ibanez, R. Rabadan and
A.M. Uranga, Nucl. Phys. B \textbf{542} (1999) 112 [arXiv:hep-th/9808139].

\end{thebibliography}
\end{document}